\documentclass{v16nufact}
\pdfoutput=1

\def\be{\begin{equation}}
\def\ee{\end{equation}}
\def\bea{\begin{eqnarray}}
\def\eea{\end{eqnarray}}

\newcommand{\Photo}{}

\usepackage[backend=bibtex,natbib=true, sorting=none, sortcites]{biblatex}
\usepackage{amsmath, amssymb, slashed, siunitx}
\usepackage{subfigure}

\begin{document}
\vspace*{4cm}
\title{HEAVY NEUTRINO SEARCHES FROM MeV TO TeV\\\textsc{Proceedings of the NuFact 2016 conference}}

\author{ E. GRAVERINI }

\address{Physik Institut der Universit\"at Z\"urich\\Winterthurerstrasse 190,
	8057 Z\"urich, Switzerland}

\maketitle\abstracts{
The Standard Model (SM) describes particle physics with great precision. However, it does not account for the generation of neutrino masses, whose nature we do not understand. Both a Dirac and a Majorana mass term could intervene, leading to the existence of heavy partners of the SM neutrinos, presumably more massive and nearly sterile. For suitable choices of parameters, heavy neutrinos can also provide dark matter candidates, and generate the observed baryon asymmetry of the universe. Heavy neutrinos can be searched for at beam dump facilities such as the proposed SHiP experiment if their mass is of the order of few \si{\giga\electronvolt}, or at high energy lepton colliders, such as the Future $e^+e^-$ Circular Collider, FCC-ee, presently under study at CERN, for higher masses. This contribution presents a review of the sensitivities for heavy neutrino searches at SHiP and FCC-ee.
}

\section{Introduction}

The data from experiments with solar, atmospheric, reactor and accelerator neutrinos provide compelling evidence for the existence of a 3-neutrinos mixing process in vacuum, caused by nonzero neutrino masses, which is in contrast with the Standard Model (SM) predictions. Experiments also show that the three flavours of neutrinos (antineutrinos) $\nu_\ell$, with $\ell = e, \mu, \tau$, are always produced in a left-handed $\nu_L$ (right-handed $\bar{\nu}_R$) helicity state~\cite{Agashe:2014kda}. At present there is no experimental evidence for the existence of right-handed (left-handed) neutrinos $\nu_R$ (antineutrinos $\bar{\nu}_L$). Therefore, if such states exist, they must be interacting very weakly with SM particles, and are therefore also called ``sterile'' neutrinos.

\begin{figure}[thbp]
	\centering
	\subfigure[In the $\nu$MSM, three right-handed counterparts $N_{1,2,3}$ are added to the particle content of the SM~\cite{PP}.]{\includegraphics[width=0.47\textwidth]{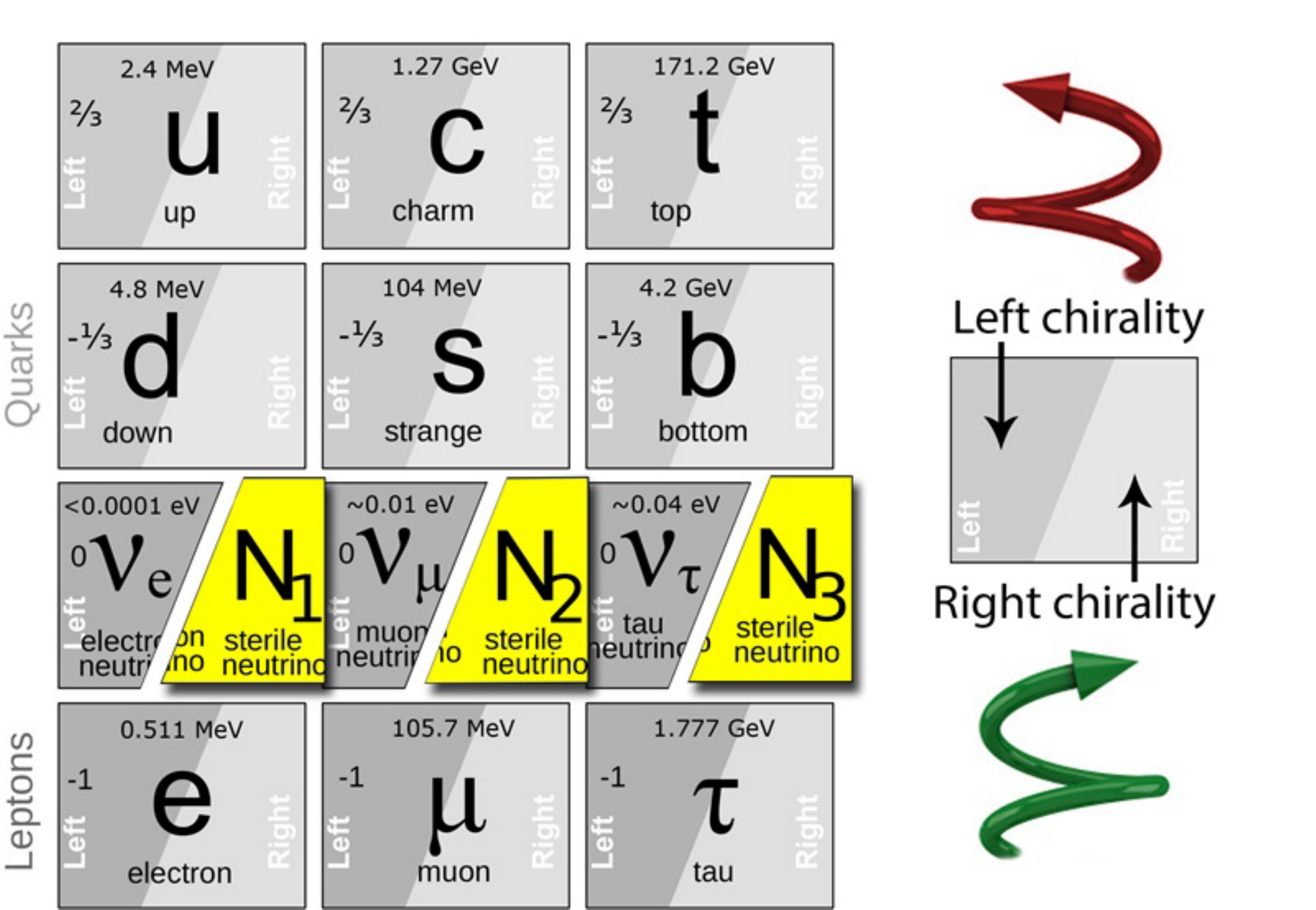}\label{numsm}}\hfill
	\subfigure[Decay of an heavy neutrino through mixing with a SM neutrino~\cite{Graverini:2214085}.]{\includegraphics[width=0.47\textwidth]{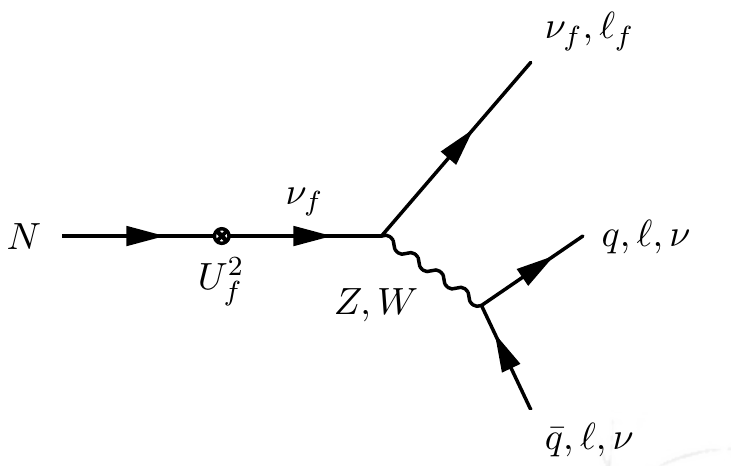}\label{hnldecay}}
	\caption{Sketch of the particle content of the $\nu$MSM, and the decay of heavy neutrinos.}
\end{figure}

There is not a unique way to extend the Standard Model in order to account for neutrino masses. Both a Dirac mass term, or a Majorana mass term, would imply the existence of right-handed neutrinos; the cohexistence of these two terms would lead to the type I seesaw mechanism. One of the most popular theories accounting for neutrino masses, the $\nu$MSM~\cite{nuMSM}, sketched in \figurename~\ref{numsm}, suggests Majorana masses ranging from sub-\si{\mega\electronvolt} to \si{\tera\electronvolt}, and Dirac masses smaller or similar to the electron mass. Three right-handed neutrino states complement the left-handed SM neutrinos, addressing all the known shortcomings of the SM. Depending on the masses and couplings of the right-handed neutrinos, the lightest one would provide a Dark Matter candidate in the keV region, while the two others, in the \si{\mega\electronvolt}-\si{\giga\electronvolt} range, would provide mass to the SM neutrinos through the seesaw mechanism, and generate asymmetry between matter and antimatter through leptogenesis.

\section{The $\nu$MSM}

The $\nu$MSM complements the particle content of the SM by adding three right-handed neutrinos, also referred to as HNLs (Heavy Neutral Leptons). The SM lagrangian extends to:
\begin{align}
	\mathcal{L} = \mathcal{L}_{SM} + \bar{N_i}i\slashed\partial N_i - f_{i\alpha}\Phi\bar{N_i}L_{\alpha} - \frac{M_i}{2}\bar{N_i}^c N_i + h.c.
\end{align}
where $\Phi$ and $L_\alpha (\alpha = e, \mu, \tau)$ are respectively the Higgs and lepton doublets, $f$ is a matrix of Yukawa couplings and $M_i$ is a Majorana mass term. The $N_i$ fields represent right-handed neutrinos.
One of the new states, $N_1$, is a very long lived dark matter candidate, with a lifetime possibly exceeding the age of the Universe, and a mass in the $\mathcal{O}(\si{\kilo\electronvolt})$ range~\cite{Boyarsky:2009ix}.
The other two states have very similar mass $m_N$, being almost degenerate, with $m_N$ in the \si{\mega\electronvolt}-\si{\giga\electronvolt} range. They would cause the baryon-antibaryon asymmetry of the Universe through a process of leptogenesis enabled by the lepton number violating Majorana mass term~\cite{nuMSM2}. The two heavier states would also produce the observed neutrino masses through the type I seesaw mechanism, first introduced in the context of Grand Unified Theories~\cite{Minkowski:1977sc,GellMann:1980vs}.

\subsection{Heavy neutrino phenomenology}

\begin{figure}[thbp]
	\centering
	\hfill
	\subfigure[Production in charm meson decays.]{\includegraphics[width=0.35\textwidth]{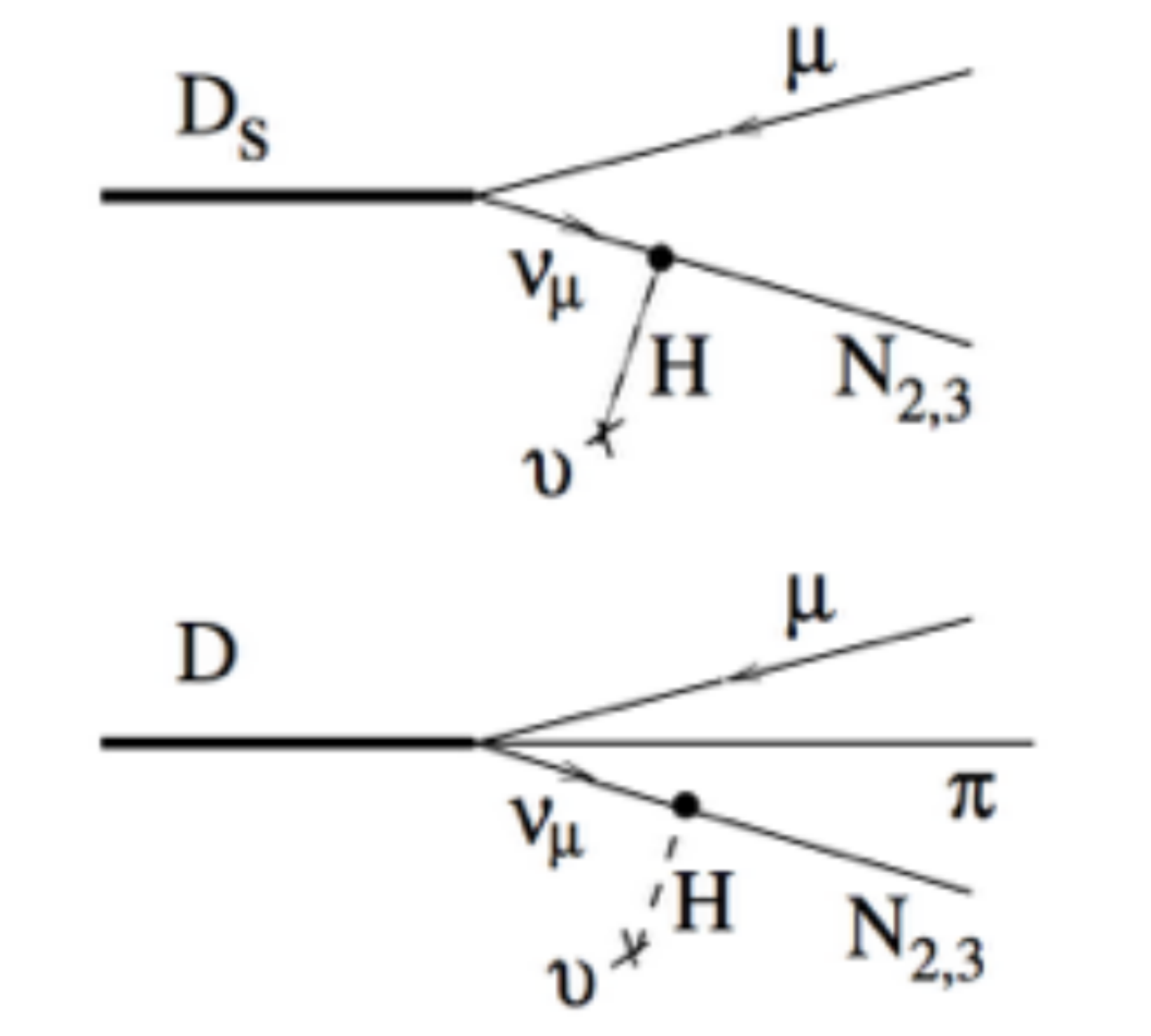}\label{dsd}}\hfill
	\subfigure[Production in $W^\pm, Z$ decays.]{\includegraphics[width=0.39\textwidth]{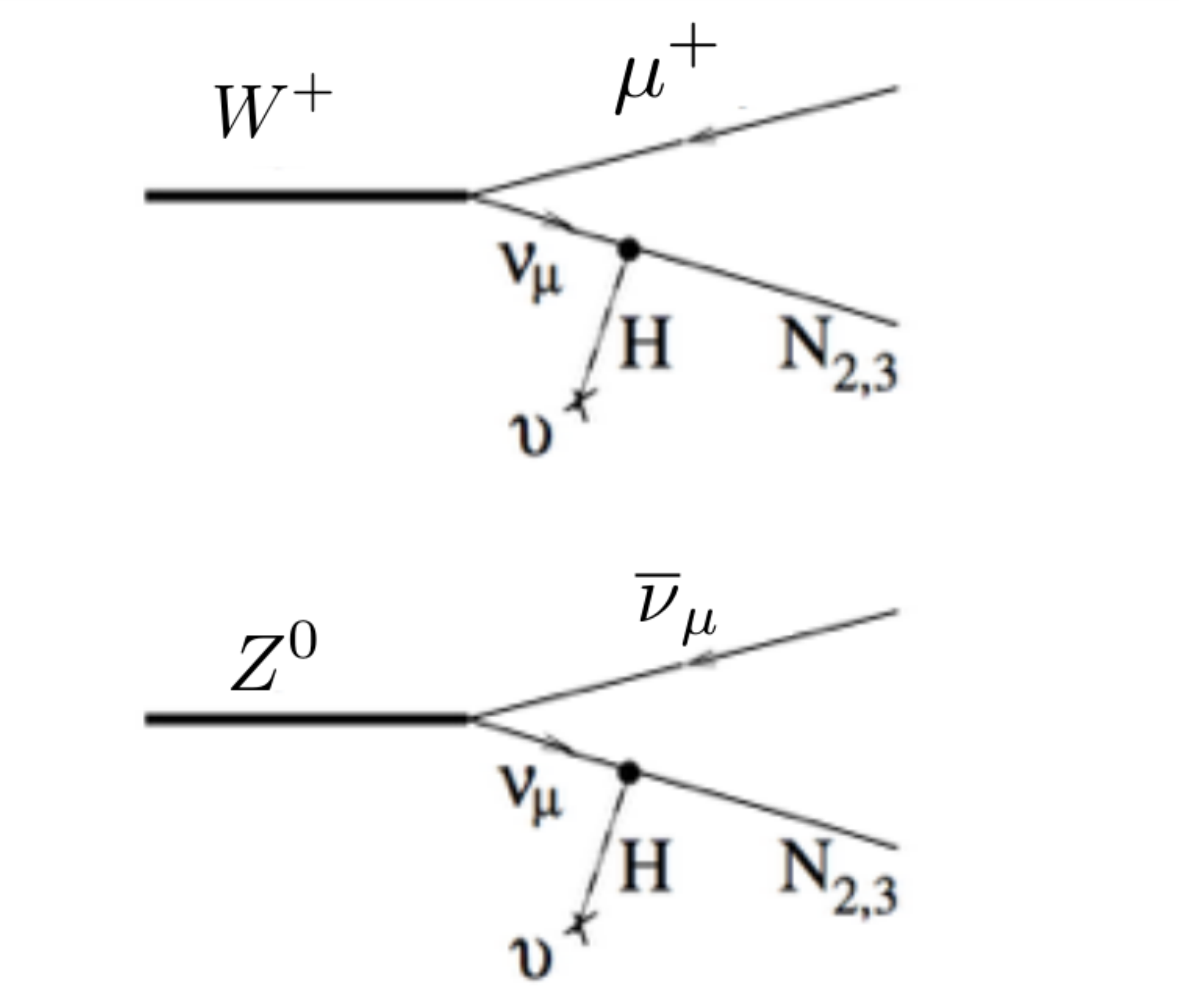}\label{wz}}\hfill
	\caption{Production of HNLs the decays of mesons and gauge bosons.}
\end{figure}

The phenomenology of sterile neutrino production is thoroughly described in~\cite{Gronau:1984ct,Gorbunov:2007ak,Atre:2009rg}. HNLs can be produced in decays where a SM neutrino is replaced by an HNL through kinetic mixing. They then decay to SM particles by mixing again with a SM neutrino. These now massive neutrino states can decay to a variety of final states through the emission of a $W^\pm$ or $Z^0$ boson (see \figurename~\ref{hnldecay}).
Branching ratios for the production and decay of HNLs can be obtained from those for the light neutrino channels as shown in~\cite{Gorbunov:2007ak}. In particular, heavy neutrinos with mass up to a few \si{\giga\electronvolt} can be produced in semileptonic decays of $\pi, K, D$ and $B$ mesons (\figurename~\ref{dsd}). Heavier right-handed neutrinos can emerge in decays of the $W^\pm$ and $Z$ bosons where one neutrino is replaced by a HNL (\figurename~\ref{wz}).

HNLs up to a few GeV in mass can decay to a meson-lepton pair, or to three leptons ($h\ell$, $h^0\nu$, $\ell\ell'\nu$, $3\nu$ final states). If the HNL mass is substantially larger than the QCD scale $\Lambda_{QCD}$, however, the two emitted quarks will tend to hadronize separately, producing two jets in the detector, accompanied by either a charged lepton or a neutrino ($jj\ell$, $jj\nu$). The three-lepton final state remains as likely as for low mass HNLs.

% Now describe the available parameter space
The parameter space of the $\nu$MSM is bound on all sides (\figurename~\ref{img:sensitivities}).
Low mass HNLs would participate in the Big Bang Nucleosynthesis (BBN) and modify the observed relative abundances of light nuclei. Very low couplings, depending on the HNL mass, cannot explain the observed SM neutrino mass differences with the seesaw mechanism. Finally, the necessity to justify the baryon asymmetry of the Universe (BAU) limits the parameter space from above: if the mixing is too large, HNLs would be in thermal equilibrium during the expansion of the Universe, and thus fail to produce baryogenesis through oscillation.
For HNL masses approaching $m_W$, the rate of interactions is enhanced due to the now kinematically allowed decay channel $N_{2,3} \to \ell W$, leading to stronger constraints on the mixing parameter~\cite{Shaposhnikov:2008pf, my-ichep}.

\section{Prospects for low mass searches}

\begin{figure}[thbp]
	\centering
	\subfigure[The SHiP proposed experiment~\cite{TP}]{\includegraphics[width=.58\textwidth]{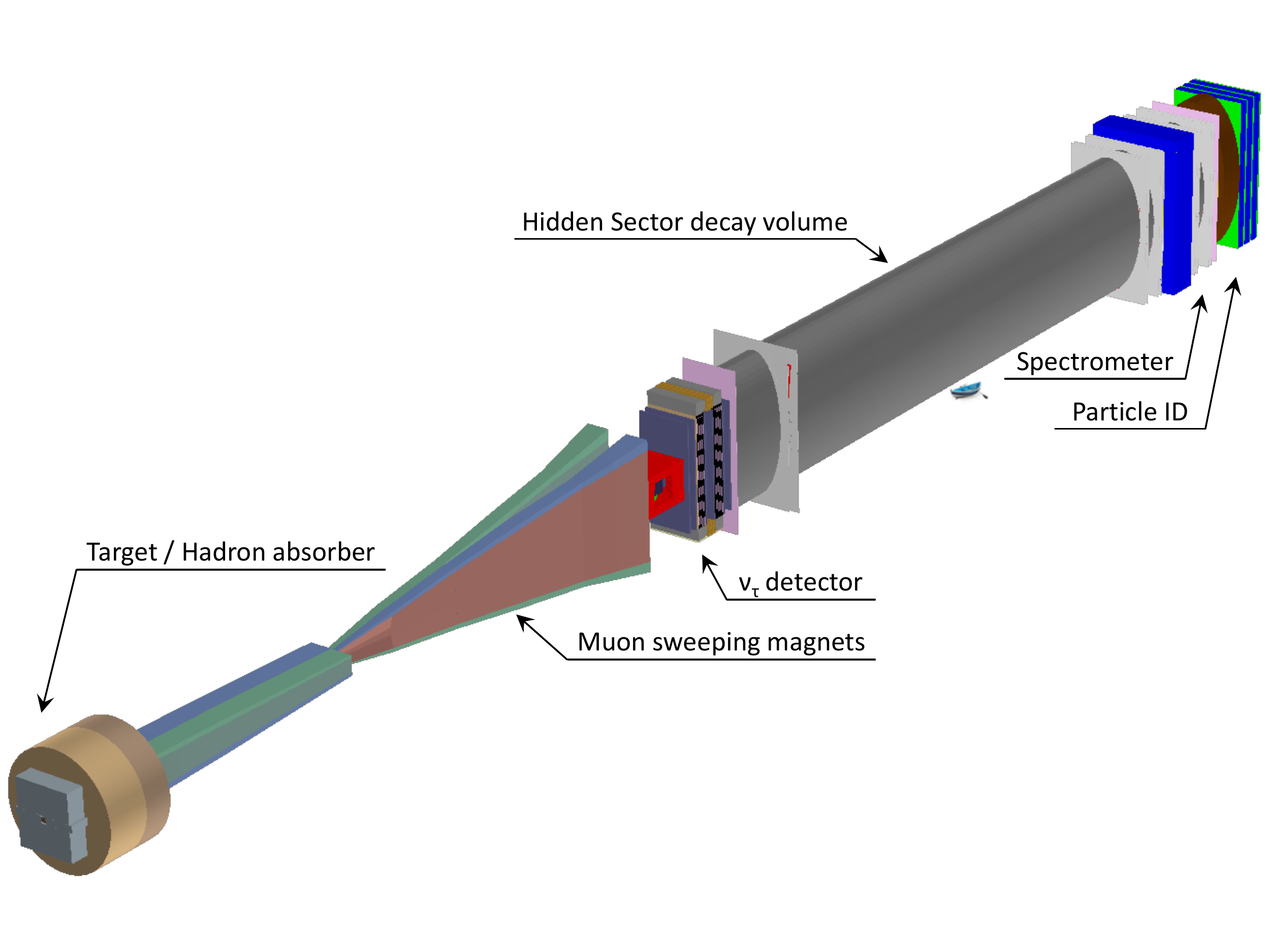}\label{img:ship_layout}}
	\hfill
	\subfigure[The FCC proposal~\cite{fcc-design}]{\includegraphics[width=.38\textwidth]{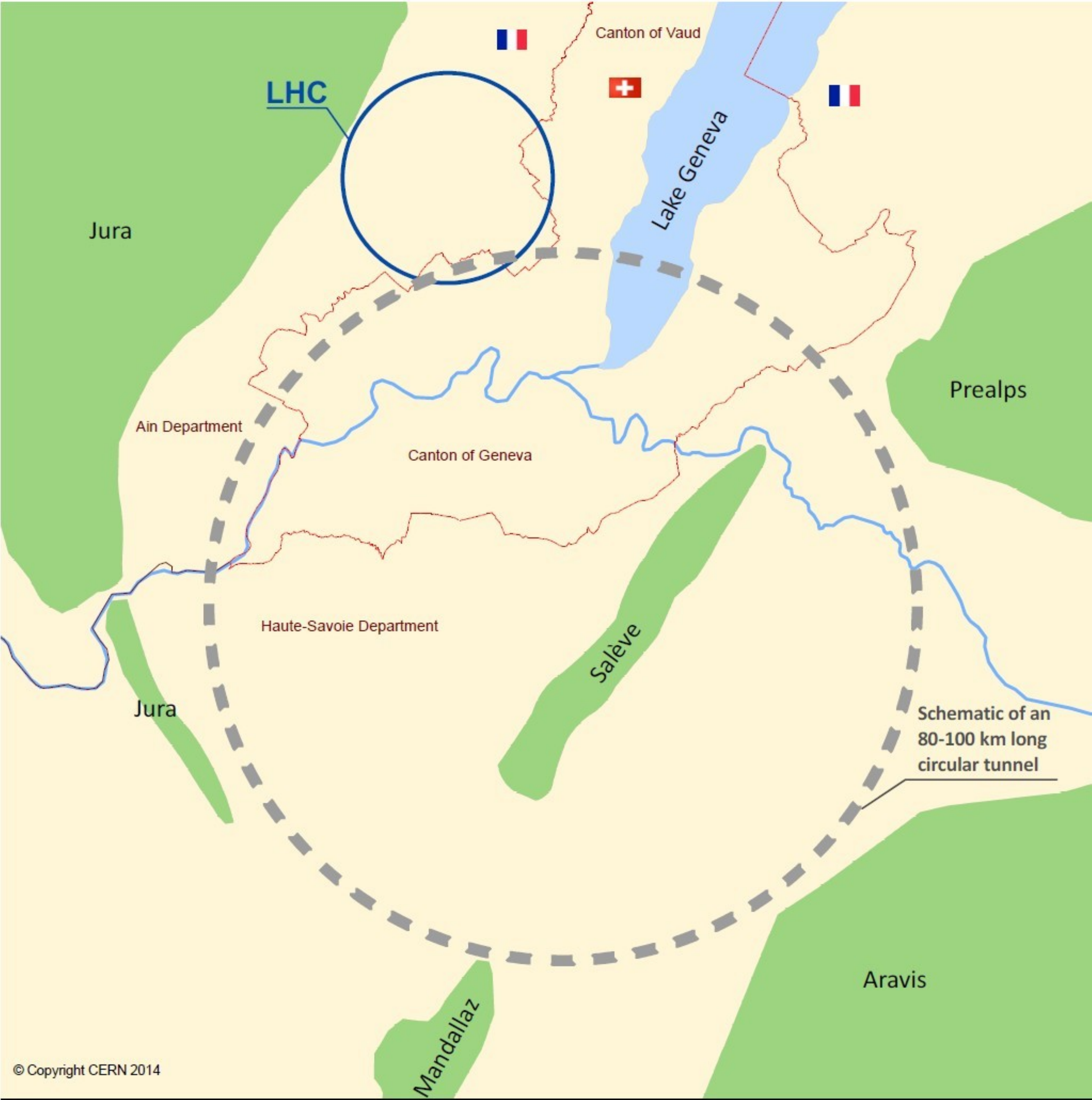}\label{img:fcc}}
	\caption{Proposed facilities that could probe for heavy neutrinos in the near or medium term future.}\label{img:future_exp}
\end{figure}

SHiP is a newly proposed general-purpose fixed target facility at CERN, with the aim of looking for very weakly interacting long living particles. The 400 GeV/$c$ SPS proton beam will be stopped in a heavy target, designed to maximize the production of heavy mesons while reducing that of neutrinos and muons to a minimum. An iron hadron stopper will absorb secondary hadrons and residual non-interacting protons. Muons emerging from the beam dump will be deflected from the detector area by a system of magnets of alternate polarity.
The hidden sector detector will search for evidence of hidden particles decaying in a large fiducial volume, contained in a 62~m long vacuum vessel with elliptical cross-section of 5$\times$10~m$^2$. A straw-tubes magnetic spectrometer is placed at the end of the vacuum tank. The particle identification system comprises hadron and electromagnetic calorimeters, and a muon detector.
Conservative studies show that a level of vacuum of $10^{-6}$ atm reduces the background due to SM neutrino interactions to less than 0.1 events in 5 years. In addition, the decay volume is preceded by two background taggers, surrounded by a layer of liquid scintillator for the whole length of the vessel, and followed by a high precision timing detector. This allows to control backgrounds due to random combination of tracks and to upstream interactions, effectively making SHiP a ``zero-background'' experiment~\cite{TP}. The SHiP experimental layout is shown in \figurename~\ref{img:ship_layout}.

Past experiments have set important constraints on the parameter space for HNLs. The most significant limits below the charm mass were obtained in the fixed target experiments PS191~\cite{Bernardi:1987ek,Vannucci:2008zz}, CHARM~\cite{Bergsma:1985is} and NuTeV~\cite{Vaitaitis:1999wq}.
% Searches in $B$ and $Z^0$ decays, and in electron beam dump experiments, are in general sensitive but cover a different region of the parameter space.
Table~\ref{tab:intro_pastexp} lists the relevant parameters of the above three experiments, in comparison with those planned for SHiP~\cite{TP}.
For each one of these experiments, no signal candidates were found. Regions of the HNL parameter space were excluded, down to couplings $U^2$ as low as $10^{-6}$ (CHARM, NuTeV) or even $10^{-8}$ (PS191), limited to masses below 450~MeV due to the low energy of the PS beam.

\begin{table}[!htbp]
	\begin{center}
		\begin{tabular}{l  r r r r }\hline
			Experiment & PS191 & NuTeV & CHARM & SHiP\\
			\hline
			Proton energy (GeV)& 19.2 & 800 & 400 & 400\\
			Protons on target ($\times 10^{19}$)& 0.86 & 0.25 & 0.24 & 20\\
			Decay volume (m$^3$)& 360 & 1100 & 315& 1780\\  
			Decay volume pressure (bar)& 1 (He) & 1 (He) & 1 (air) & $10^{-6}$ (air)\\
			Distance to target (m) & 128 & 1400& 480& 80-90\\
			Off beam axis (mrad)& 40 & 0 & 10 & 0\\
			%$\#$ background events& $<1$& $0.57\pm0.15$& ? &$<10^{-2}$\\
			\hline
		\end{tabular}
	\end{center}
	\caption{Comparison of the experimental conditions for HNL search experiments~\cite{TP}.}\label{tab:intro_pastexp}
\end{table}

As shown in \figurename~\ref{ship}, SHiP will greatly improve the sensitivity of the previous experiments using the production of heavy hadrons at the SPS.
In particular, 8$\times$10$^{17}$ $D$~mesons
and 3$\times$10$^{15}$ $\tau$ leptons 
are expected in about 5 years of nominal SPS operation. Beauty hadrons will also contribute to the physics sensitivity.
% between the beauty hadron and 
%the charm hadron masses. 
Below the beauty hadron mass, the SHiP experiment will be able to exceed the sensitivity of previous experiments for HNLs
by several orders of magnitude. HNL couplings could be probed close to the ultimate seesaw limit.
%The sensitivities of other existing or planned experiments to hidden sector particles are more than an order of magnitude lower compared to SHiP, even under the experimentally challenging
%assumption of zero background~\cite{TP}. 

\section{Prospects for collider searches}

Heavy right-handed neutrinos can also be searched for at high luminosity lepton colliders, such as the Future $e^+e^-$ Circular Collider (FCC-ee), sketched in \figurename~\ref{img:fcc}, currently being studied at CERN~\cite{fcc-design}. The machine would fit in a 100~km tunnel and would be able to operate at centre-of-mass energies in the 90-350~GeV range.
Luminosity studies show that FCC-ee, operated at the $Z$ resonance, could produce $10^{12}$ to $10^{13}$ $Z$ bosons per year, and therefore allow to investigate extremely rare decays~\cite{PhysRevSTAB.17.041004, Koratzinos:2013ncw, Gomez-Ceballos:2013zzn}.

A review of possible methods to perform HNL searches at future $e^+e^-$ colliders is given in~\cite{my-ichep}. Hints of the existence of sterile neutrinos can be found in the discrepancy between the measured number of neutrino families -- the ratio of the $Z$ invisible width to its leptonic decay width -- and that of the SM lepton flavours. The former, $N_\nu = 2.9840 \pm 0.0082$~\cite{ALEPH:2005ab}, appears to be about two standard deviations lower than 3, and such decifit could be compatible with the presence of sterile neutrinos.
However, for small mixing angles between sterile and active neutrinos, as those predicted by all models trying to explain the matter-antimatter asymmetry, the most efficient way to look for sterile neutrinos at a high-energy lepton collider is to operate it as a $Z$ factory~\cite{my-ichep}.

\begin{figure}[thbp]
	\centering
	\subfigure[Normal hierarchy]{\includegraphics[width=.48\textwidth]{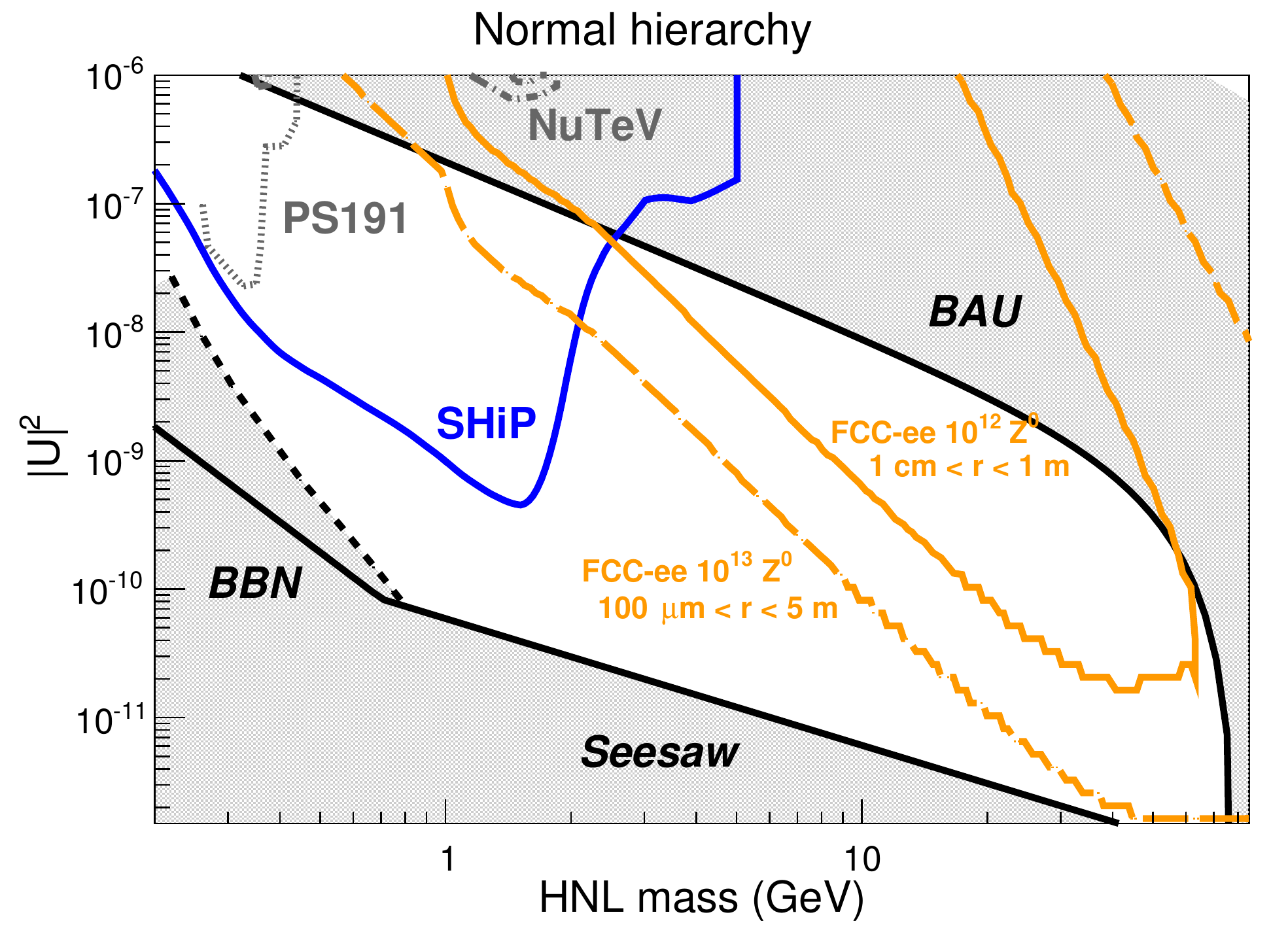}\label{ship}}
	\hfill
	\subfigure[Inverted hierarchy]{\includegraphics[width=.48\textwidth]{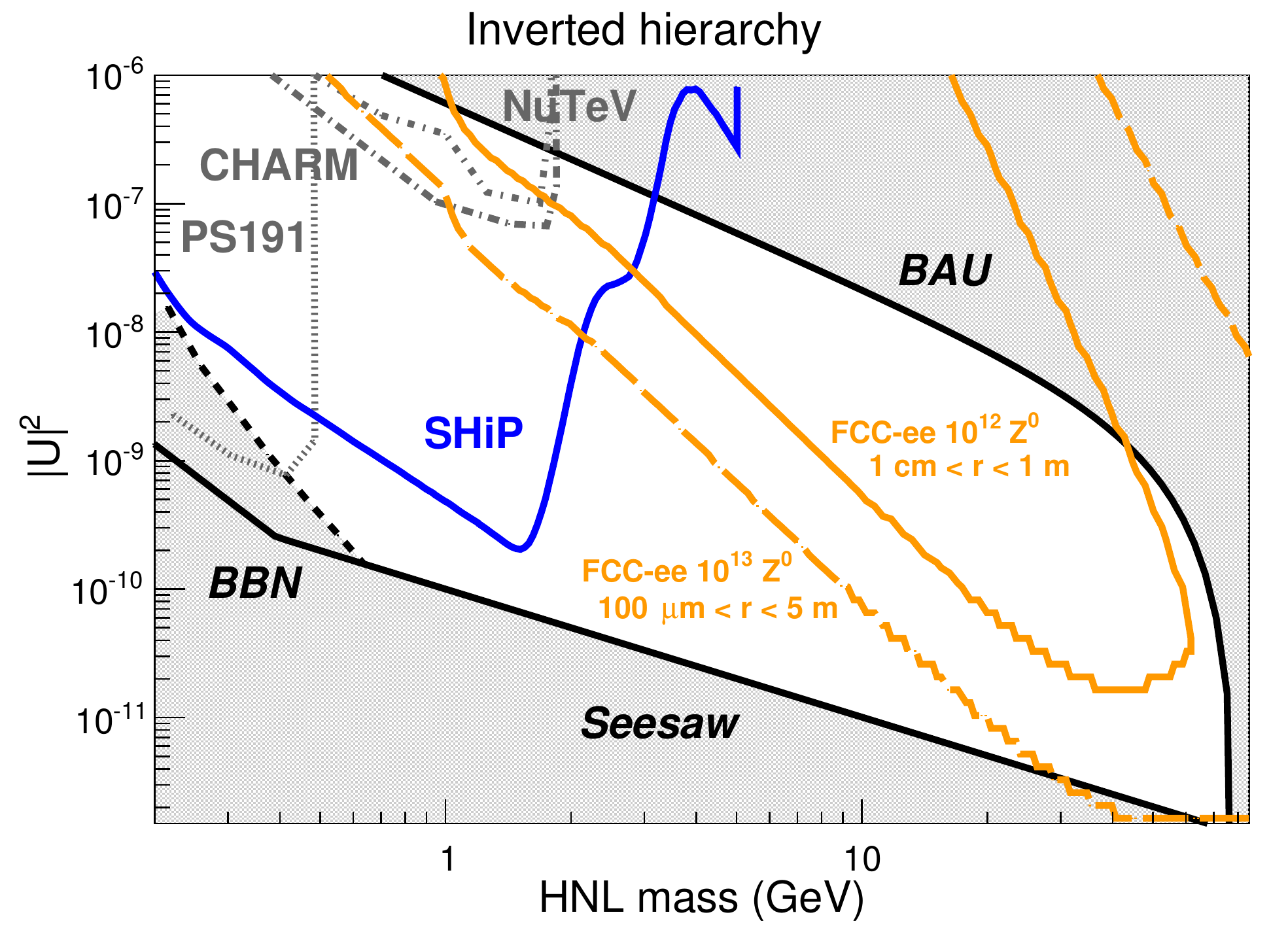}\label{fcc}}
	\caption{Physics reach in the HNL parameter space for SHiP and two realistic FCC-ee configurations and for two $\nu$MSM scenarios: normal hierarchy of SM neutrino masses and coupling to the muon flavour dominating, and inverted hierarchy with coupling to the electron flavour dominating, respectively. Previous searches are shown as dotted lines. Greyed-out areas represent the cosmological boundaries of the scenario~\cite{Graverini:2214085, my-ichep}.}\label{img:sensitivities}
\end{figure}

HNLs can be produced in $Z \to \nu\bar{\nu}$ decays with a SM neutrino kinematically mixing to an HNL, therefore producing $Z \to \nu N$. At very small couplings, the lifetime of the HNL becomes substantial, giving the possibility to suppress background arising from $W^*W^*$, $Z^*Z^*$ and $Z^*\gamma^*$ processes with the requirement of a displaced secondary vertex~\cite{my-ichep}.

\figurename~\ref{img:sensitivities} analyses the sensitivities of SHiP and of an hypotetical FCC-ee experiment in the parameter space of the $\nu$MSM, for two realistic FCC-ee configurations. The minimum and maximum displacements of the secondary vertex in FCC-ee depend on the characteristics of the tracking detectors of the experiment. For the first (second) FCC-ee configuration, an inner tracker with resolutions of 100~$\mu$m (1~cm) and an outer tracker with diameter of 1~m (5~m) were considered. The production of $10^{12}$ ($10^{13}$) $Z$ bosons is assumed.

The SHiP experiment will be able to scan a large part of the parameter space below the $B$ meson mass. In addition, the results shown in \figurename~\ref{img:sensitivities} show that heavier HNLs can be searched for at a future $Z$ factory. The synergy between a large fixed-target experiment at the SPS and a future $Z$ factory would allow to explore most of the $\nu$MSM parameter space.

%\section{Current limits and future prospects}
\printbibliography

\end{document}